\documentclass[prl,twocolumn,showpacs,superscriptaddress]{revtex4}
\usepackage{amsmath}
\usepackage{amssymb}
\usepackage{graphicx}
\begin{document}

\title{ Giant viscosity enhancement in a spin-polarized Fermi liquid}
\author{H. Akimoto}
\affiliation{Microkelvin Laboratory, NHMFL, and Physics Department, University of Florida, Gainesville, FL 32611}
\affiliation{Physics Department, University of Massachusetts, Amherst, MA 01003}
\affiliation{Nano-Science Joint Laboratory, RIKEN, Wako, Saitama 351-0198, Japan}
\author{J. S. Xia}
\affiliation{Microkelvin Laboratory, NHMFL, and Physics Department, University of Florida, Gainesville, FL 32611}
\author{D. Candela}
\affiliation{Physics Department, University of Massachusetts, Amherst, MA 01003}
\author{W. J. Mullin}
\affiliation{Physics Department, University of Massachusetts, Amherst, MA 01003}
\author{E. D. Adams}
\affiliation{Microkelvin Laboratory, NHMFL, and Physics Department, University of Florida, Gainesville, FL 32611}
\author{N. S. Sullivan}
\affiliation{Microkelvin Laboratory, NHMFL, and Physics Department, University of Florida, Gainesville, FL 32611}

\begin{abstract}
	The viscosity is measured for a Fermi liquid, a dilute $^3$He-$^4$He mixture, under extremely high magnetic field/temperature conditions ($B \leq 14.8$~T, $T \geq 1.5$~mK).
	The spin splitting energy $\mu B$ is substantially greater than the Fermi energy $k_B T_F$; as a consequence the polarization tends to unity and $s$-wave quasiparticle scattering is suppressed for $T \ll T_F$.
	Using a novel composite vibrating-wire viscometer an enhancement of the viscosity is observed by a factor of more than 500 over its low-field value.
	Good agreement is found between the measured viscosity and theoretical predictions based upon a $t$-matrix formalism.
\end{abstract}

\date{18 April 2007}
\maketitle

	Spin polarization provides a powerful tool for investigating the quantum states of Fermi liquids, such as degenerate electrons, ultracold atoms, and $^3$He-$^4$He mixtures.
	Polarization affects the transport coefficients of the system due to Fermi statistics \cite{meyerovich90}.
	In low magnetic fields, the polarization tends to a value less than unity as the temperature is lowered below the Fermi temperature $T_F$ (normal Pauli paramagnetism).
	However, if the magnetic field exceeds a critical value, the Fermi energy is smaller than the spin splitting and the polarization tends exponentially to unity as the temperature is lowered \cite{mullin86}.
	Here we report an experiment that attains this regime for the first time in a $^3$He-$^4$He mixture, by employing extremely large field/temperature ratios $B/T$ up to 5900~T/K.
	As a consequence the spin polarization attains values greater than 99\%, and the viscosity we measure for the liquid increases by a factor of more than 500 over its low-field value.

\begin{figure}[t]
\centering
\includegraphics[width=0.8 \linewidth]{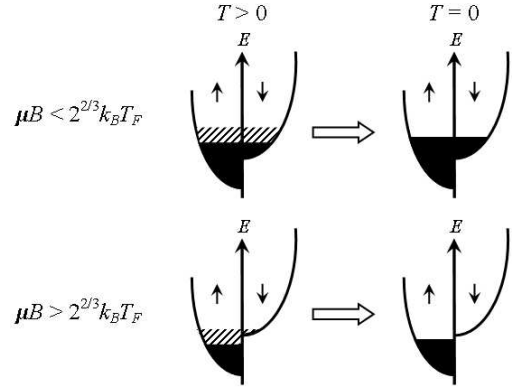}
\caption{\label{tffig}
	Schematic illustration of mechanism for attaining extremely high spin polarizations in degenerate Fermi liquids.
	The parabolic curves show the densities of states for spin-up and spin-down, offset from one another by the spin splitting $\mu B$.
	Well below the Fermi level states are fully occupied (black), while within $k_B T$ of the Fermi level states are partially occupied (shaded).
	For small to moderate fields (top row), the polarization remains less than unity as $T\rightarrow 0$.
	Conversely, for sufficiently large fields (bottom row), the polarization tends exponentially to unity as $T\rightarrow 0$.
}\end{figure}

	The mechanism for reaching extremely high spin polarizations is illustrated in Fig.~\ref{tffig}.
	To reach this regime the spin splitting $\mu B$ must exceed $2^{2/3}$ times the Fermi energy $k_B T_F$, where the factor of $2^{2/3}$ arises because the favored-spin Fermi level increases when all of the spins are aligned ($\mu =$~magnetic moment of spins, $B =$~applied magnetic field, $k_B =$~Boltzmann constant).
	In $^3$He-$^4$He mixtures as employed here, the superfluid $^4$He solvent serves as a ``mechanical vacuum'' that prevents the $^3$He atoms from condensing at any temperature, thus providing one of Nature's best examples of a weakly-interacting Fermi gas~\cite{baym91}.
	The Fermi temperature is set by the concentration $x_3$ of $^3$He atoms, $T_F = (2.66\mbox{~K})x_3^{2/3}$, and the spin splitting energy in temperature units is $\mu B/k_B = (1.56\mbox{~mK/T})B$.
	Therefore, very dilute $^3$He-$^4$He mixtures at very low temperatures in very high magnetic fields are required to reach the high-polarization regime.

	Spin polarization affects the viscosity (transport of momentum) and other transport coefficients mainly via its effect upon quasiparticle-quasiparticle scattering \cite{meyerovich90}.
	Due to the exclusion principle, only anti-parallel pairs of spins can scatter in $s$-wave orbital states, and the fraction of anti-parallel pairs decreases with increasing spin polarization.
	As a result the mean free path becomes longer and the viscosity and thermal conductivity grow as the polarization is increased.
	In a fully polarized system, $p$-wave scattering would dominate. 
	Previous experiments measured an increase of viscosity with spin polarization in $^3$He-$^4$He mixtures, initially via the attenuation of second sound~\cite{owersbradley88} and more recently via the direct damping of a vibrating wire~\cite{candela91},\cite{owersbradley98}.
	However these earlier experiments did not meet the condition $\mu B >2^{2/3}k_B T_F$ hence the spin polarizations and viscosity enhancements were much smaller than those reported here.
	Exploration of the high-polarization regime provides important tests of transport theory for highly-polarized systems.
	In particular we test whether a semi-phenomenological $t$-matrix formalism that successfully describes Fermi liquids at low to moderate polarization can be extended to these extreme conditions.

	To achieve the extremely high $B/T$ ratios required for this experiment we used a cryostat combining a powerful nuclear demagnetization refrigerator with a 15~T NMR magnet for the sample region.
	The nuclear stage contains 92~mol of copper (effectively 46 mol in 8~Tesla) along with 5~mol of PrNi$_5$ \cite{xia04}.
	The temperature was monitored using a melting pressure thermometer in the low magnetic field region~\cite{rusby02}, supplemented by a Kapton capacitance thermometer inside the sample cell.
	The capacitance thermometer was calibrated in high $B/T$ conditions in an earlier experiment at much higher $x_3 = 3.8\%$~\cite{akimoto03}, in which the viscosity was nearly field independent and could be used as a high $B/T$ thermometer.
	The $^3$He-$^4$He sample was mixed to give a nominal $^3$He concentration $x_3=200$~ppm, but $x_3$ in the sample cell is less than this value due to trapping of $^3$He atoms in the sample fill line.
	By fitting the sample magnetization (measured by pulsed NMR) to the magnetization of a free Fermi gas and to the magnetization measured in an earlier, higher-$x_3$ experiment, the $^3$He concentration in the sample cell was determined to be $x_3 = 150 \pm 15$~ppm.

	The largest viscosity enhancement in $^3$He-$^4$He reported prior to the present work was a factor of 3.5, at 75\% spin polarization~\cite{owersbradley98}.
	That work used a viscometer made from fine (25~$\mu$m diameter) wire, to achieve sensitivity for the very low viscosity sample ($\eta \sim \mathcal{O}(10^{-6}$~Pa-s)).
	Consequently a significant slip correction was required, as the $^3$He mean free path (MFP) in the sample became larger than the wire diameter at the lowest temperatures and highest fields.
	To measure the viscosity in the much longer MFP conditions of the present experiment a new viscometer design was required, which combines a weak restoring force (for sensitivity) with a large surface area (to minimize slip).
	Our experimental cell contained both a conventional fine-wire viscometer and a viscometer with this new composite design.

	The conventional vibrating-wire viscometer (\mbox{VW-A}) was a semicircular loop of fine wire (nominal 0.0015~in. dia. un-annealed manganin-290 wire, California Fine Wire Co., Grover City, CA). 
	The wire diameter measured with a scanning electron microscope was 36.7~$\mu$m.
	The legs of \mbox{VW-A} were 5~mm apart and were fixed in place with Sycast 1266 epoxy.
	This viscometer had a resonant frequency in vacuum of 3438~Hz.

\begin{figure}[t]
\centering
\includegraphics[width=1.0 \linewidth]{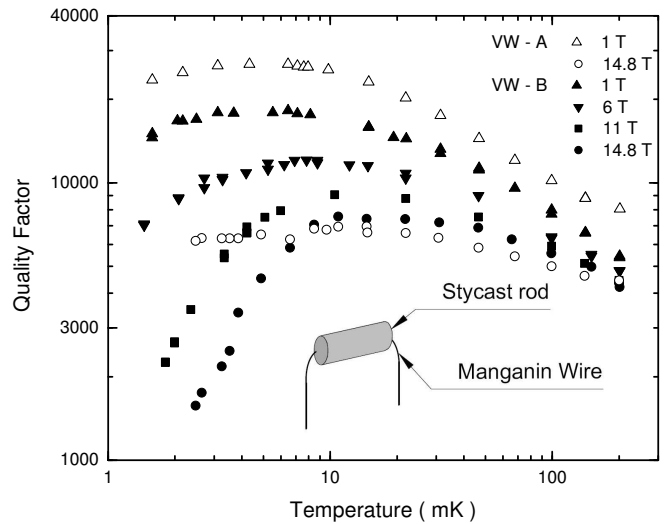}
\caption{\label{qfig}
	Measured quality factors for the two vibrating-wire viscometers in a $^3$He-$^4$He mixture with $^3$He concentration $x_3 = 150$~ppm.
	The conventional viscometer \mbox{VW-A} loses sensitivity at high fields due to slip.
	The composite viscometer \mbox{VW-B}, sketched in the inset, incorporates an enlarged section to reduce slip effects.
}\end{figure}

	The other viscometer (\mbox{VW-B}) consisted of an 0.84~mm diameter epoxy rod glued to the central portion of a 36.7~$\mu$m-diameter manganin wire loop as shown in the inset of Fig.~\ref{qfig}.
	The overall height of \mbox{VW-B} was 8~mm, and the distance between the legs was 6~mm.
	For this viscometer the epoxy rod was sufficiently large to interact hydrodynamically with the quasiparticle gas at the lowest temperatures and highest fields used.
	In comparison with the conventional viscometer, \mbox{VW-B} had a very low resonant frequency (96.1~Hz in vacuum).

	Figure~\ref{qfig} shows the quality factors measured for the two viscometers in the $^3$He-$^4$He sample, over a range of fields and temperatures (a preliminary report with some of this unanalyzed quality-factor data was made in Ref.~\cite{akimoto02}).
	At high fields and temperatures below 10~mK, the conventional viscometer loses all sensitivity due to severe MFP effects.
	Conversely the composite viscometer VW-B retains sensitivity for the lowest temperatures and highest fields and requires only a modest slip correction.

	To find the viscosity we used the analysis of Stokes~\cite{stokes01} for an oscillating cylinder supplemented by a slip correction as calculated by H{\o}jgaard Jensen et al. \cite{hojgaardjensen80},
\begin{equation}
	k + ik' =1+[(k_0-1+ik_0')^{-1} - iM^2 \beta ]^{-1}.
\end{equation}
	Here $k_0, k_0'$ are the dimensionless forces on the wire~\cite{stokes01} in the absence of slip while $k, k'$ are the same quantities corrected for slip, calculated as functions of $M=r/\sqrt{2}\delta$.
	The radius of the wire is $r$ and the viscous penetration depth is $\delta =(2 \eta /\rho_n \omega)^{(1/2)}$, where $\rho_n$ is the normal-fluid density proportional to $x_3$.
	The contribution of the $^3$He quasiparticle gas to the inverse of the viscometer quality factor is $Q_H^{-1} = (\rho_n/\rho_w) k'$, allowing $k'$ (and hence the viscosity $\eta$) to be measured.
	Recently Bowley and Owers-Bradley \cite{bowley04} have improved upon the slip calculation of H{\o}jgaard Jensen et al. \cite{hojgaardjensen80} by including the effect of the curvature of the wire surface, and this theory has been tested experimentally by Perisanu and Vermeulen~\cite{perisanu06}.
	However, this new slip calculation has not yet been extended to cover the very long MFP conditions of the present experiment.

	Following Refs.~\cite{carless83} and~\cite{guenalt83} we take slip correction parameter
\begin{equation}
	\beta = 0.579 ~ ( \ell / r )(1 + \alpha \ell/r)/(1 + \ell/r),
\end{equation}
where $\ell$ is the $^3$He mean free path.
	The parameter $\alpha$ is fixed by the ballistic-limit ($\ell/r \rightarrow \infty$) damping of the wire.
	Carless et al.~\cite{carless83} estimated $\alpha=2.46$ for superfluid $^3$He-B, and Gu\'{e}nault et al.~\cite{guenalt83} fit their data on $^3$He-$^4$He with $\alpha=2.3$.
	Our data fit $\alpha=1.0\pm 0.2$, so we have used this value of $\alpha$ to correct our data for slip.
	The mean free path $\ell$ is calculated from the viscosity measured with VW-B, iterating the calculation as necessary until self-consistent values of $\ell$ and $\eta$ are found.
	For $B=1$~T we find $\ell = 4.7$~$\mu$m at 100~mK and $\ell = 22$~$\mu$m at 1~mK.
	Therefore at this low field, the slip effect correction appears only as a modest correction for \mbox{VW-A} at temperatures below 3~mK.
	At lower temperatures and in higher magnetic fields, the mean free path becomes much larger, $\ell \approx 3$~mm at 2.5~mK in 14.8~Tesla.
	Nevertheless it remains comparable to the 0.84~mm diameter of the epoxy rod on viscometer \mbox{VW-B}, making accurate viscosity measurements possible.

\begin{figure}[t]
\centering
\includegraphics[width=1.0 \linewidth]{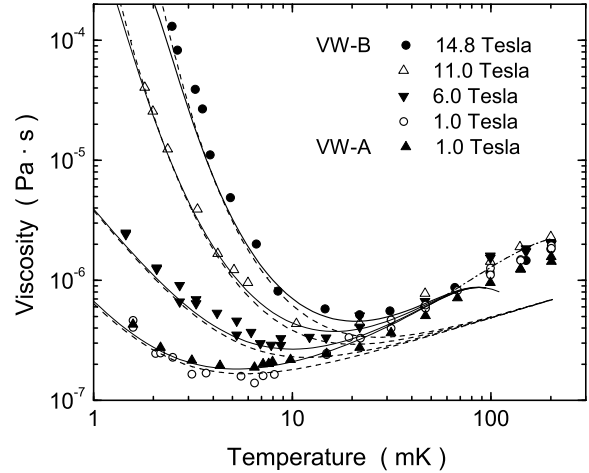}
\caption{\label{etahighfig}
	Data points:~Viscosity of the $^3$He-$^4$He sample over a wide range of fields.
	At $B=1.0$~T the slip corrections are modest for both viscometers, and the viscosity data from \mbox{VW-A} and \mbox{VW-B} are in good agreement.
	At higher fields only the composite viscometer \mbox{VW-B} is used.
	Solid curve:~Calculation of the viscosity using a quadratic approximation to the $t$-matrix element $V(q)=V_0[1-(q/q_0)^2]$.
	The downturn of this curve for $T>70$~mK is an artifact of truncating $V(q)$ at the quadratic term.
	This is illustrated by the dash-dot curve, which shows a high-temperature calculation using an untruncated $V(q)$ following Ref.~\cite{lhuillier82}.
	Dashed curve:~Calculation using $V(q)=V_0$, which corresponds to an $s$-wave approximation for quasiparticle scattering.
}\end{figure}

\begin{figure}[t]
\centering
\includegraphics[width=1.0 \linewidth]{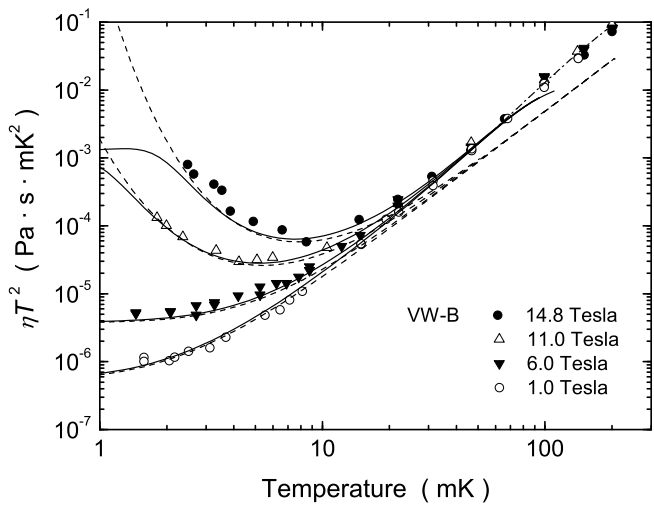}
\caption{\label{etat2fig}
Viscosity multiplied by temperature squared, using the same notations for data and calculations as in Fig.~\ref{etahighfig}.
	For lower fields $B < 7.6$~T, $\eta T^2 \rightarrow $~constant as $T \rightarrow 0$ is observed.
	For fields above this critical value $\eta T^2$ has a minimum and then increases as $T \rightarrow 0$.
	At temperatures lower than the range of the data, the $s$-wave calculation of $\eta T^2$ (dashed curves) diverges, while the quadratic $V(q)$ calculation (solid curves) tends to a constant due to $p$-wave scattering. 
}\end{figure}

	Figure \ref{etahighfig} shows the viscosity of the $^3$He-$^4$He sample derived from the viscometer quality factors (with slip correction) as detailed in above.
	The two viscometers \mbox{VW-A} and \mbox{VW-B} are in good agreement at the relatively low field $B=1$~T.
	This validates the operation of the composite viscometer \mbox{VW-B}, which then can be used to determine the sample viscosity at extremely high-field/low-temperature conditions.
	As can be seen in Fig.~\ref{etahighfig}, the viscosity measured at $B=14.8$~T, $T=2.5$~mK is more than 500 times greater than the viscosity at low field and the same temperature.

	The distinction between high and low field regimes is clear when the viscosity data are multipied by temperature squared, Fig.~\ref{etat2fig}.
	For lower fields satisfying $\mu B<2^{2/3}k_B T_F$ (corresponding to $B < 7.6$~T in the present experiment), normal Fermi-liquid behavior $\eta T^2 \rightarrow$~constant as $T \rightarrow 0$ is observed.
	For fields above this critical value $\eta T^2$ has a minimum and then increases for $T \ll T_F$, due to the exponential suppression of $s$-wave scattering.

	These data provide a new test for the theory of transport in degenerate, highly polarized Fermi liquids.
	The interaction between two $^3$He quasiparticles in a $^3$He-$^4$He mixture is thought to be well-understood; it is mediated by virtual $^4$He phonons and velocity-dependent interactions should be small due to the small ratio of the $^3$He Fermi velocity ($v_F=4.3$~m/s for $x_3 = 150$~ppm) to the $^4$He sound velocity (240~m/s)~\cite{baym91}.
	Therefore, this system provides a stringent test of many-body and quantum-statistical effects upon transport.
	Although $s$-wave scattering between quasiparticles dominates at low temperatures and Fermi energies, corrections beyond $s$-wave are significant for some of the data presented here (solid \emph{vs} dashed curves in Figs.~\ref{etahighfig}-\ref{etat2fig}).

	As in early theoretical treatments of $^3$He-$^4$He solutions~\cite{baym91},\cite{bardeen67},\cite{ebner67} the $t$-matrix representing scattering of two $^3$He's is approximated by a spin and velocity independent interaction $V(q)$ depending only on momentum transfer $q$, $\langle p+q, p'-q |t| p, p' \rangle \approx V(q)$.
	The theoretical curves in Figs.~\ref{etahighfig}-\ref{etat2fig} were computed on this basis~\cite{mullin92}, using the form for $V(q)$ developed by Ebner~\cite{ebner67}.
	For the low-temperature part of the theoretical curves it was computationally necessary (and adequate) to truncate this potential to a quadratic form, $V(q) \approx V_0[1-(q/q_0))^2]$.
	The effect upon transport of adding quadratic corrections to $V_0$ has also been considered by Hampson et al.~\cite{hampson88}. 

 	In this language the $s$-wave scattering length is proportional to $V(0)=V_0$, while corrections beyond $s$-wave correspond to higher-order terms $V_0(q/q_0)^2 \dots$.
	Some work has called into question the use of an effective interaction $V(q)$, or more generally the possibility of accurate theoretical predictions beyond the $s$-wave limit~\cite{meyerovich90}.
	Hence, it is valuable to determine experimentally the domain of accuracy of calculations that attempt to calculate transport beyond $s$-wave.
	
	The theoretical formalism used here has been used successfully to fit most aspects of a recent experiment on transverse spin dynamics in $^3$He-$^4$He at much higher $x_3$~\cite{akimoto03}, as well as earlier experiments that measured modest viscosity increases at much lower polarizations~\cite{candela91}.
	As shown by the curves in Fig.~\ref{etahighfig}, the $t$-matrix calculation can fit our viscosity data over the entire range of fields and temperatures up to $70\mbox{~mK}\approx 10 T_F$ with only two input parameters $V_0$, $q_0$.
	In particular, the large viscosity enhancement at the highest fields and lowest temperatures is accurately predicted.
	The value $V_0 = -1.84\times 10^{-38}$~erg~cm{$^3$} used for the calculated curves in Figs.~\ref{etahighfig}-\ref{etat2fig} is 23\% larger than the value that was used to fit lower-field, $x_3 =630$~ppm viscosity data in Ref.~\cite{candela91}, and 31\% larger than the value originally developed by Ebner~\cite{ebner67} to fit transport data at much higher $x_3$.
	Conversely we have not adjusted $q_0/\hbar = 4.03\times 10^7$~cm$^{-1}$ from the original Ebner value.
	The apparent modest variation of $V_0$ with $^3$He concentration $x_3$, if real, might signal the existence of corrections not included in the $t$-matrix formalism.
	Clearly, however, this formalism captures the dominant behavior of both spin and momentum transport over wide ranges of polarization, $^3$He concentration, and temperature.

	This work was supported by the Science Research Program of the National High Magnetic Field Laboratory funded by the NSF-DMR-9016241 and the State of
Florida.

\bibliography{visco3}

\end{document}